\documentclass{emulateapj}
\usepackage{epsf}
\usepackage{apjfonts}
\usepackage{xspace}
\usepackage{amsmath}
\usepackage{framed} 

   \usepackage{graphicx}

\def\HI{{\rm H\,I}}
\def\HII{{\rm H\,II}}

\def\tb{\Delta T_B}
\def\tbsub{\tb^{\rm SUB}}
\def\tbigm{\tb^{\rm IGM}}

\def\dim#1{\mbox{\,#1}}

\def\hide#1{}

\begin{document}

\title{Cosmic Reionization On Computers. Mean and Fluctuating Redshifted 21 cm Signal}

\author{Alexander A.\ Kaurov\altaffilmark{1} and Nickolay Y.\ Gnedin\altaffilmark{2,1,3}}
\altaffiltext{1}{Department of Astronomy \& Astrophysics, The
  University of Chicago, Chicago, IL 60637 USA; kaurov@uchicago.edu}
\altaffiltext{2}{Particle Astrophysics Center, 
Fermi National Accelerator Laboratory, Batavia, IL 60510, USA; gnedin@fnal.gov}
\altaffiltext{3}{Kavli Institute for Cosmological Physics and Enrico
  Fermi Institute, The University of Chicago, Chicago, IL 60637 USA} 

\begin{abstract}
We explore the mean and fluctuating redshifted 21 cm signal in numerical simulations from the Cosmic Reionization On Computers (CROC) project. We find that the mean signal varies between about $\pm25\dim{mK}$. Most significantly, we find that the negative pre-reionization dip at $z\sim10-15$ only extends to $\langle\tb\rangle\sim-25\dim{mK}$, requiring substantially higher sensitivity from global signal experiments that operate in this redshift range (EDGES-II, LEDA, SCI-HI, and DARE) than has been often assumed previously. We also explore the role of dense substructure (filaments and embedded galaxies) in the formation of 21 cm power spectrum. We find that by neglecting the semi-neutral substructure inside ionized bubbles, the power spectrum can be mis-estimated by 25-50\% at scales $k\sim 0.1-1h\dim{Mpc}^{-1}$. This scale range is of a particular interest, because the upcoming 21 cm experiments (MWA, PAPER, HERA) are expected to be most sensitive within it.
\end{abstract}


\section{Introduction}
\label{sec:intro}

Redshifted 21 cm signal  from the epoch of reionization has been considered as an extremely powerful but highly futuristic probe of the spatial distribution of neutral hydrogen in the intergalactic medium (IGM) for a long time. Commonly, the redshifted 21 cm signal is split into a global, averaged over the whole sky (aka mean) signal $\langle\tb\rangle$ and the fluctuating component; the latter can be characterized  by the power-spectrum $P_k$ or by some other statistical technique (and, eventually, with direct imaging). Observations of these two components rely on very different instrumental techniques, and can be treated as effectively two different (but related) observational probes of cosmic reionization.

Time may be approaching, however, when the actual signal is finally detected: recent constraints from Murchison Widefield Array \citep[MWA; ][]{21cm:dlw14,21cm:dnw15} and Donald C.\ Backer Precision Array for Probing the Epoch of Reionization \citep[PAPER][]{21cm:pla14,21cm:jpp15,21cm:pap15,21cm:apz15}) are steadily edging towards the predicted cosmological signal. While no detection has been made so far, the current progress was deemed sufficient by NSF to start funding the next generation experiment, Hydrogen Epoch of Reionization Array (HERA), that is expected to achieve a robust detection before the end of this decade \citep{21cm:hera}.

While the observational detection was distant, simple analytical and semi-numerical techniques were sufficient for modeling the cosmological signal theoretically, and for making an order-of-magnitude estimates of the expected signal. An excellent recent review of these methods is given by \citet{21cm:pl12}, so we only mention two of them: a widely cited prediction of the mean signal by \citet{21cm:pl08} and a publicly available code 21CMFAST  \citep{21cm:mfc11} for making maps of the fluctuating 21 cm signal.

As the whole field matures and the detection of the cosmological signal appears to become increasingly likely, it is important to continuously refine and improve theoretical modeling, making sure that all crucial physical processes are adequately included. The recognition of this need resulted in a field-wide effort of developing ``next generation'' modeling capabilities, both with full cosmological numerical simulations and with advanced semi-numerical techniques. Projects like ``DRAGONS'' \citep{newrei:dwm15}, ``Renaissance Simulations'' \citep{newrei:owx15}, ``Cosmic Dawn'' \citep{newrei:dawn}, ``Emma'' \citep{newrei:emma}  make a certainly incomplete list.

In this paper, we rely on numerical simulations from the Cosmic Reionization On Computers (CROC) project \citep{ng:g14,ng:gk14} as our theoretical model. Of course, for a given simulation to serve as a test of a simple model, the physical realism of the simulation  should be sufficiently high. CROC simulations are useful in this regard, since they include most of the physical effects thought to be important for modeling cosmic reionization (such as star formation and feedback, spatially-inhomogeneous and time-dependent radiative transfer, non-equilibrium, radiation-field-dependent ionization and cooling, etc) in a fully-coupled manner, in simulation volumes exceeding 100 comoving Mpc in size and with proper spatial resolution approaching $100\dim{pc}$.

Another useful feature of CROC simulations is the availability of simulation boxes of various sizes and resolutions, which allows one to test the dependence of physical predictions from the simulations on these numerical parameters.

In this paper we model redshifted 21cm emission with two sets of simulations: a set of three $40h^{-1}\dim{Mpc}$ runs labeled as B40.sf1.uv2.bw10 in \citet{ng:g14} and a single new $80h^{-1}\dim{Mpc}$ simulation (B80.sf1.uv15.bw10). In each simulation volume, we compute a $1024^3$ 2D$\times$1D (sky $\times$ frequency) grid of 21 cm brightness temperature at each simulation snapshot, accounting for all relevant physical effects (kinetic - spin temperatures coupling by Ly-$\alpha$ radiation and collisions with electrons and neutral atoms, redshift space distortions, line cone effects, etc) - since CROC simulations include full 3D radiative transfer at native resolution, we have access to the full spatially and timely resolved radiation spectrum at each simulation location. We then use the gridded values to compute various observational quantities, like the mean signal and the power spectrum.

\section{Mean 21 cm Signal}
\label{sec:mean}

\begin{figure}
\includegraphics[width=\hsize]{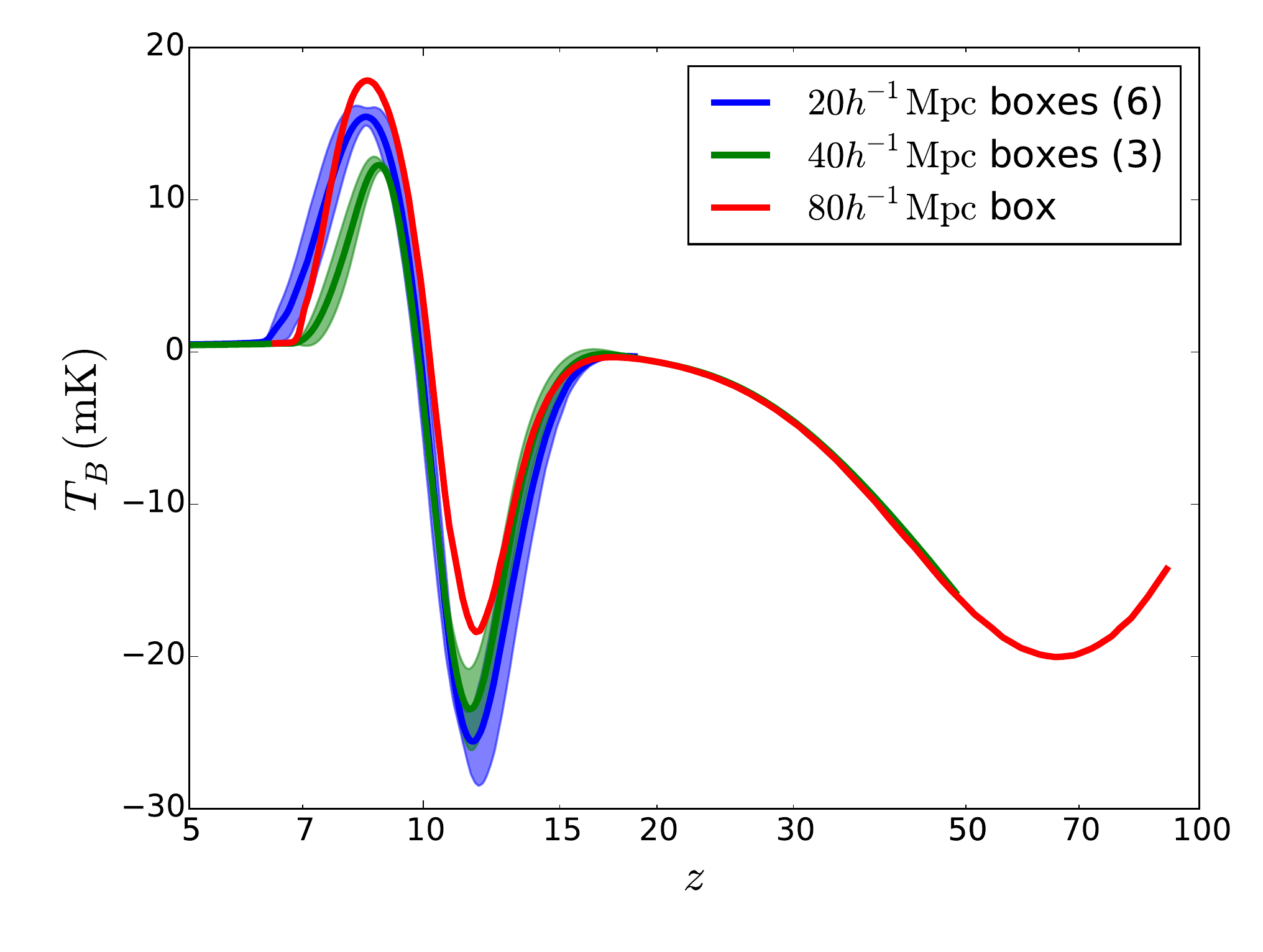}
\caption{Mean (aka ``global'') 21 cm signal as a function of redshift for three sets of CROC simulations with different box sizes (lines -  averages over all realizations, shaded bands - rms scatter between individual simulation boxes). In our fully self-consistent runs, the negative pre-reionization dip at $z\sim10-15$ reaches only to about $-25\dim{mK}$, rather then $\sim-100\dim{mK}$, as predicted by \protect\citet{21cm:pl08}.}\label{fig:meantb}
\end{figure}

The mean, or ``global'', 21 cm signal in CROC simulations is shown in Figure \ref{fig:meantb}. Our results are reasonably consistent in boxes of various sizes, although variations between simulations sets on the order of $5\dim{mK}$ remain, which should be considered as an estimate of our theoretical error. These predictions are substantially different from the most widely used model of \citet[][see also \citet{21cm:pl12}]{21cm:pl08}. Most significantly, the negative pre-reionization dip at $z\sim10-15$ only reaches down to $\tb\approx -25\dim{mK}$, rather than $-100\dim{mK}$, as predicted by \citet{21cm:pl08}. 

Following the notation by \citet{21cm:pl08} we can derive the parameters $f_\alpha$ and $f_X$ that correspond to the production efficiency of Ly$\alpha$ photons and X-rays per baryon in stars (see \citet{21cm:pl08} for the details). In our fiducial model the values are $1$ and $9$ respectively.

In fact, earlier simulations of \citet{ng:gs04} found a similarly low value for the dip, and interpreted the difference with (a similar earlier work to) \citet{21cm:pl12} as a new and dominant heating mechanism of cosmic gas before a sufficient Ly-$\alpha$ background is built up: shock heating by small-scale structure. They showed that without shock heating (or, more precisely, with the $P\,dV$ term in the energy balance equation artificially switched off), the pre-reionization dip indeed extends to $\tb\sim-100\dim{mK}$, in agreement \citet{21cm:pl08}. That claim was challenged by \citet{21cm:mo12}, who showed that shock heating due to structure formation caused only small amount of heating. However, their simulation boxes ($0.2h^{-1}\dim{Mpc}$ and $0.5h^{-1}\dim{Mpc}$) were way too small to account for the bulk of power in the velocity field (which, for Planck cosmology, peaks at a scale of about $80h^{-1}\dim{Mpc}$), and, hence, are not guaranteed to be representative for the mean universe. 

\begin{figure}
\includegraphics[width=\hsize]{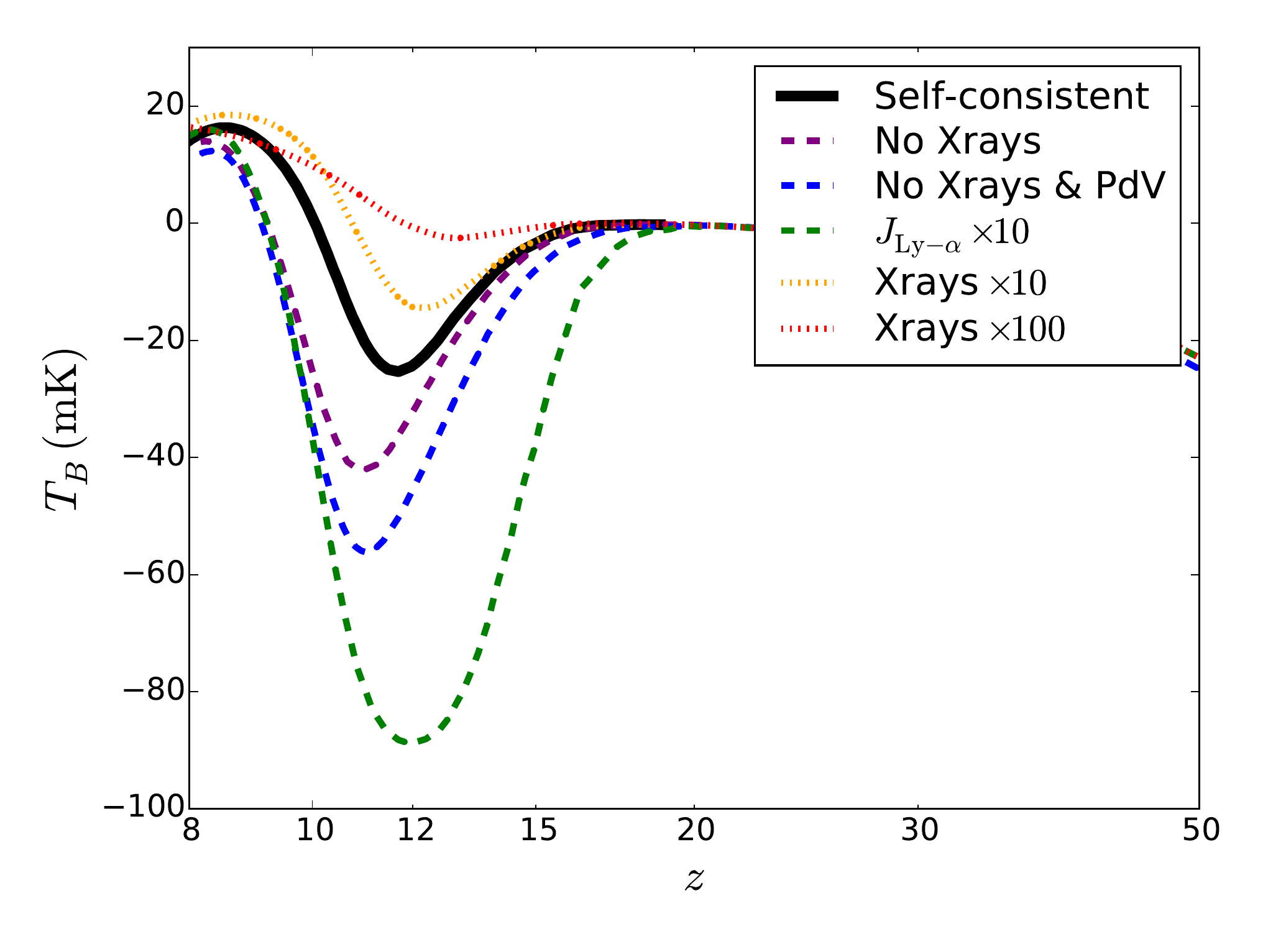}
\caption{Mean 21 cm signal as a function of redshift for simulations with varied physics. In qualitative agreement with \protect\citet{21cm:mo12}, the contribution of the $P\,dV$ term is modest, and the depth of the negative pre-reionization dip at $z\sim10-15$ is modulated by a combination of X-ray heating and spin temperature coupling by Ly-$\alpha$ radiation, as has been emphasized by \protect\citet{21cm:pl08}.}\label{fig:meanvar}
\end{figure}

Our full simulation results are consistent with \citet{ng:gs04}, so their actual results are confirmed; however, their interpretation can and should be challenged. To that end, we repeated and extended the tests performed by \citet{ng:gs04}. Such extensive parameter study would be unfeasible for any except our smallest, $20h^{-1}\dim{Mpc}$ boxes, and only for one simulation rather than all six. Therefore, we selected one realization (box A) that has its mean signal most closely matched to the average over all 6 boxes, and ran several modifications of that simulation with varied physical assumptions. The results of such exploration are shown in Figure \ref{fig:meanvar}. 

A single physical component with the largest effect on the pre-reionization dip is the intensity of cosmic Ly-$\alpha$ background: setting it to zero removes any mean signal altogether (we do not show such a line, since it would be trivial), while increasing it 10-fold deepens the dip to the values ($-100\dim{mK}$) quoted by \protect\citet{21cm:pl08}. Removing X-rays (all radiation above 4 Rydbergs) extends the dip by almost a factor of 2, while excluding in addition the $P\,dV$ term from the energy equation makes only a modest further increase in the dip depth. Hence, we do not confirm the claim made by \citet{ng:gs04} about the dominance of the $P\,dV$ term, although their claim is not completely wrong (just quantitatively off by a factor of 2-3), since the $P\,dV$ term does matter somewhat. There are many reasons that can produce such a discrepancy, from different cosmology and star formation/feedback model to the lower quality of the numerical hydro scheme used by \citet{ng:gs04}; it does not seem fruitful to explore this discrepancy further since CROC simulations are far superior to the \citet{ng:gs04} ones in any imaginable respect.

Conversely, increasing the strength of the X-ray emission from our modeled stars (which is dominated by Wolfe-Rayer stars and, hence, may be deemed somewhat uncertain) by a factor of 10 substantially reduces the depth of the dip, and pumping the X-ray emission by a factor of 100 erases the dip altogether. Thus, we also confirm the conclusions of \citet{21cm:pl08}, who emphasized the critical role of the interplay between the X-ray heating of the gas and the Ly-$\alpha$ coupling between the gas temperature and the CMB. The large difference between our prediction for the dip depth ($\sim-25\dim{mK}$) and the value of $\sim-100\dim{mK}$ quoted by \citet{21cm:pl08} is in the assumptions on the strengths of these two contributions. Naturally, we would like to think that our results are more realistic, since CROC simulations treat all relevant physical processes in a ``self-consistent'' manner, in a sense that the full spectral shape of stellar radiation, from UV to X-rays, is taken from Starburst99 models \citep{misc:lsgd99}, and, therefore, the relationships between Ly-$\alpha$, ionizing, and X-ray backgrounds are as realistic as Starburst99 models are. Since CROC simulations match essentially all existing observational constraints on reionization, including observed galaxy UV luminosity functions at all redshifts $z\ga6$, they reproduce correct star formation histories of individual galactic halos. They, thus, provide a complete model of the reionization process and our results cannot be adjusted or tuned to modify the depth of the dip by more than a modest amount allowed by the current uncertainties on various observational constraints.

Of course, no model is complete, and some of the physical processes that CROC simulations do not include may provide important contributions to X-ray or Ly-$\alpha$ backgrounds. Unfortunately, most plausible enhancements to our models, such as X-rays from stellar binaries or miniquasars, would only reduce the depth of the pre-reionization dip even further. The only reasonable possibility to increase the depth would be if Starburst99 models were underestimating stellar UV and Ly-$\alpha$ emission significantly (by a factor of several). At present, that seems unlikely, though.

Our results may bring bad news to several planned global signal experiments that rely on the deep pre-reionization dip to detect the signal, such as low-frequency EDGES-II \citep{21cm:br10}, LEDA \citep{21cm:gb12}, and SCI-HI \citep{21cm:vnj14}, as their should aim at a factor of 5 higher sensitivity to be certain to detect the signal. On the other hand, for a more sensitive experiment such as DARE \citep{21cm:blb12}, there will be enough signal-to-noise to clearly confirm or rule out our predictions. We also confirm the expected reionization signal of about $20\dim{mK}$ for global experiments that focus on $z<10$ redshift range, like the high-frequency window of EDGES-II,  BIGHORNS \citep{21cm:stw15}, or SARAS \citep{21cm:psr13,21cm:pss15}.

Finally, we have also explored the question of the role of shock heating, in order to resolve the apparent conflict between \citet{ng:gs04} and \citet{21cm:mo12}. Since, as we showed above, we find the contribution of the $P\,dV$ term to be modest, the question by itself is not particularly relevant. Nevertheless, we performed several simulations with smaller box sizes, all the way to $0.5h^{-1}\dim{Mpc}$, the size of the largest of \citet{21cm:mo12} boxes. One challenge with such tests is that the box size is too small to be even qualitatively representative volume, and such boxes would have very different reionization history (and boxes below a few Mpc would fail to reionize themselves at all). Thus, in order to maintain consistency, we imposed on such small boxes the Ly-$\alpha$ background taken from our $20h^{-1}\dim{Mpc}$ box A. With this setup, we find negligible dependence on the box size, and, hence, confirm \citet{21cm:mo12} conclusions (we do not show this in a figure, as it would be trivial, all lines roughly coinciding with each other). The interpretation of \citet{ng:gs04} - that the main role of the $P\,dV$ term is in shock heating - is, therefore, incorrect (shock heating would be strongly dependent on the box size, since, as we mentioned above, most of the velocity power is on scales around $80h^{-1}\dim{Mpc}$). Rather, the contribution of the $P\,dV$ term seems to be more involved, a combination of adiabatic heating/cooling in the over/under-dense gas and the local variation in the strength of the Ly-$\alpha$ background.

\section{Role of substructure}
\label{sec:substructure}

The dense substructure, namely filaments and embedded in them galaxies, may remain partially neutral even after the surrounding IGM is fully ionized due to the higher recombination rate \citep{Miralda-Escude2000}. These neutral patches occupy small volume but are very dense and, hence, highly biased; therefore, they may significantly contribute to the 21 cm power spectrum. Since the substructure is largely resolved in CROC simulations \citep{ng:kg15}, we can estimate the significance of its contribution to the 21 cm signal.

\begin{figure*}
\includegraphics[scale=0.8]{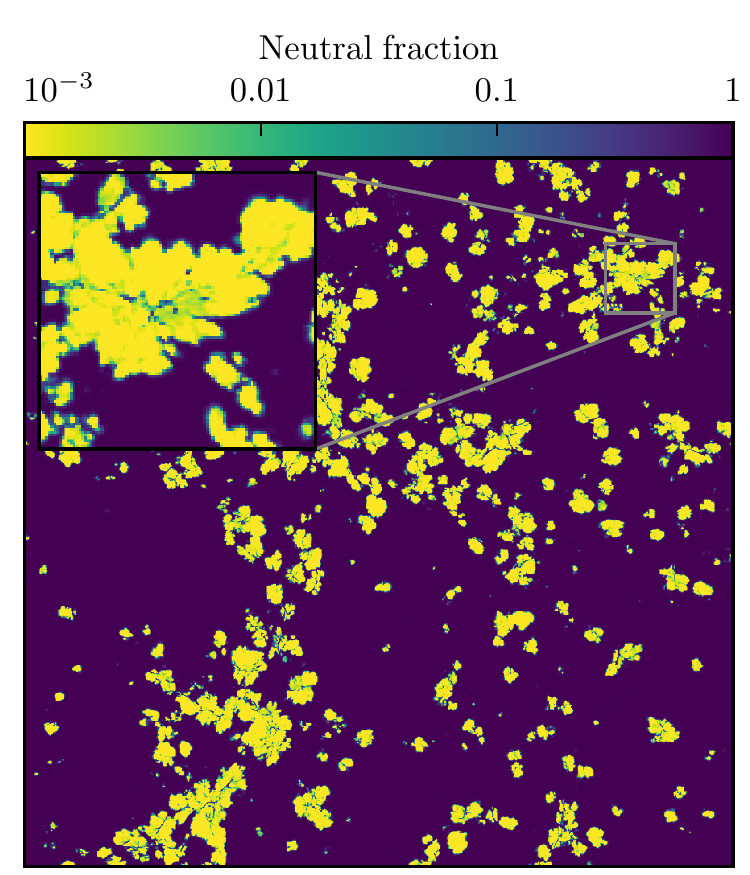}
\includegraphics[scale=0.8]{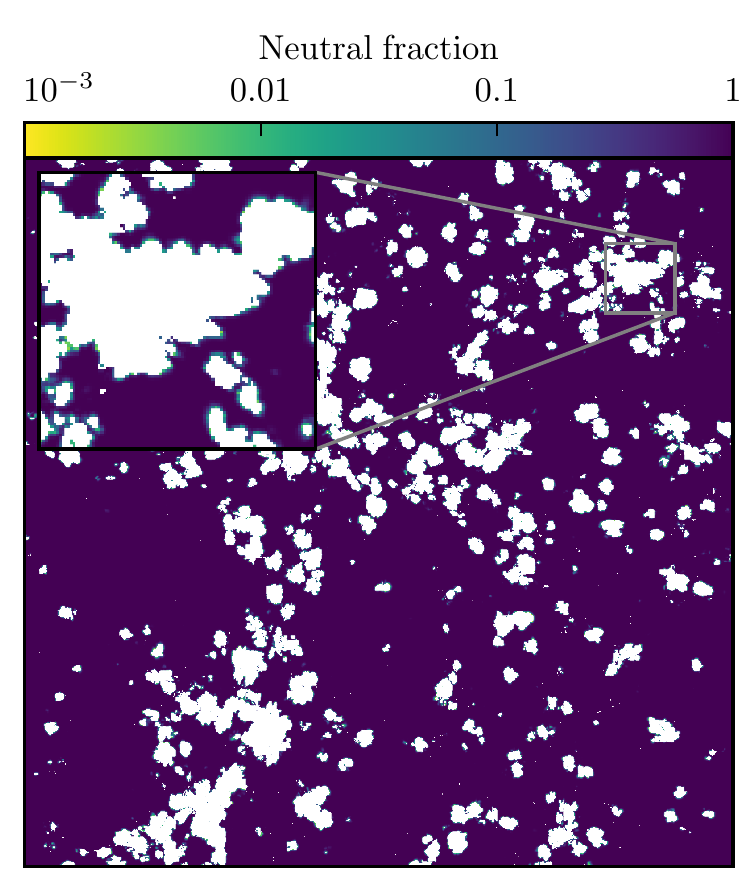}
\includegraphics[scale=0.8]{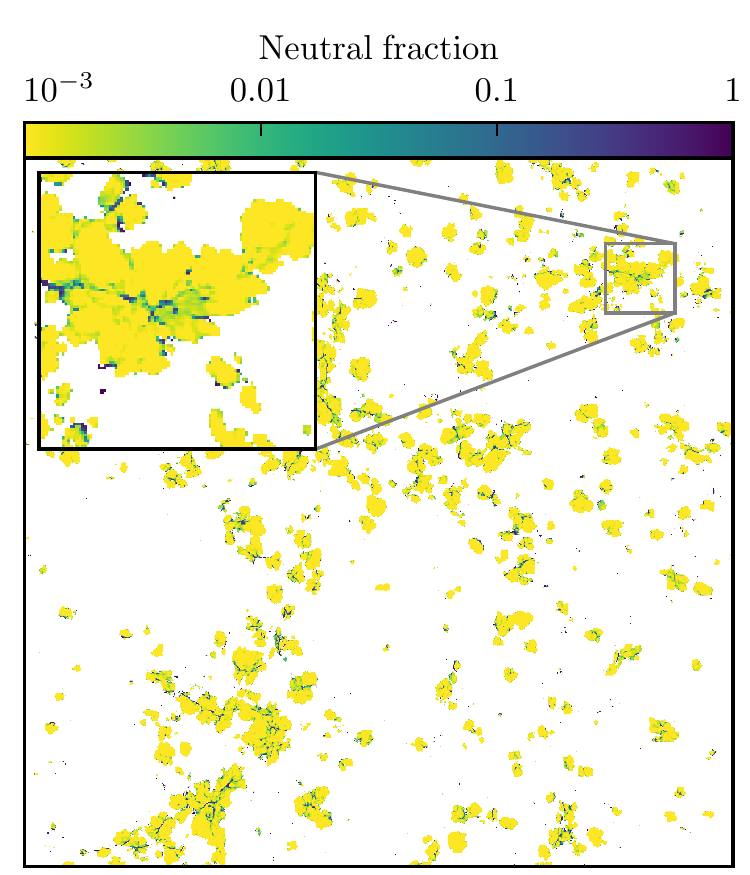}
\caption{Spatial separation of the simulation snapshot (left panel) at redshift 8.0 ($x_\mathrm{HI}=0.82$) into a ``neutral IGM'' component (middle panel) and ``ionized regions including semi-neutral substructure'' (right panel).}\label{fig:slice}
\end{figure*}

In order to test the contribution from the substructure, we spatially separate the 21 cm brightness temperature field $\tb$ into ``ionized bubbles with semi-neutral filaments'' $\tbsub$, and ``neutral IGM'' $\tbigm$. For this analysis we use our largest $80h^{-1}\dim{Mpc}$ run. The separation is performed using the ionization parameter phase space that was introduced in \citet{ng:kg15}, and the details are presented in Appendix \ref{app:a}. In Figure \ref{fig:slice} the result of such spatial separation is shown. The regions colored in white are assumed to contribute zero signal.

The $\tbigm$ field is interesting because it mimics the commonly used analytical models that do not account for the substructure. By studying the effect of adding $\tbsub$ to $\tbigm$, we will see how the power spectrum simulated by the analytical approach systematically diverges from the one calculated using the full numerical simulations with hydrodynamics and radiation transfer.

\begin{figure*}
\includegraphics[scale=0.9]{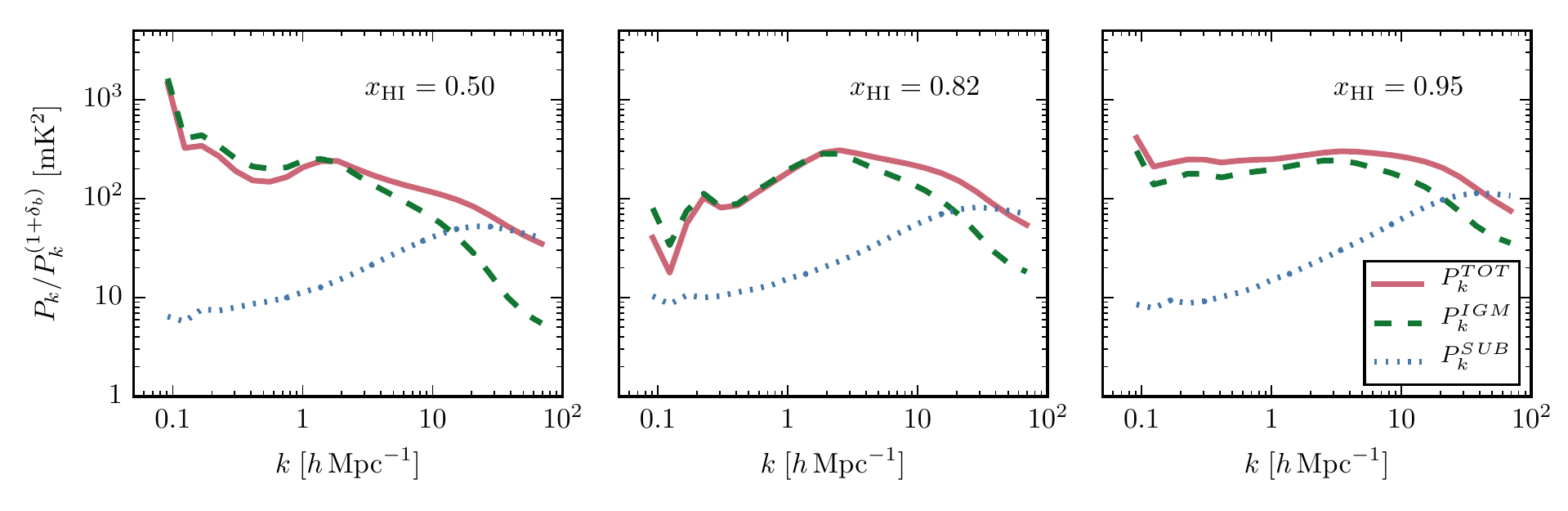}
\caption{21 cm power spectra of separate components normalized to the baryon overdensity power spectrum at different global ionization fractions. \textit{Solid red} curves correspond to the full observable emission; \textit{dashed green} lines show the component without semi-neutral substructure in ionized regions; finally, \textit{dotted blue} lines mark the semi-neutral substructure within ionized regions only. }\label{fig:ps_components}
\end{figure*}

We directly compute the 21 cm brightness temperature power spectrum of each component separately. The result is plotted in Figure \ref{fig:ps_components}. Immediately we see that $\tbsub$ and $\tbigm$ power spectra do not simply add up, because they are not independent quantities. 

\begin{figure*}
\includegraphics[scale=1]{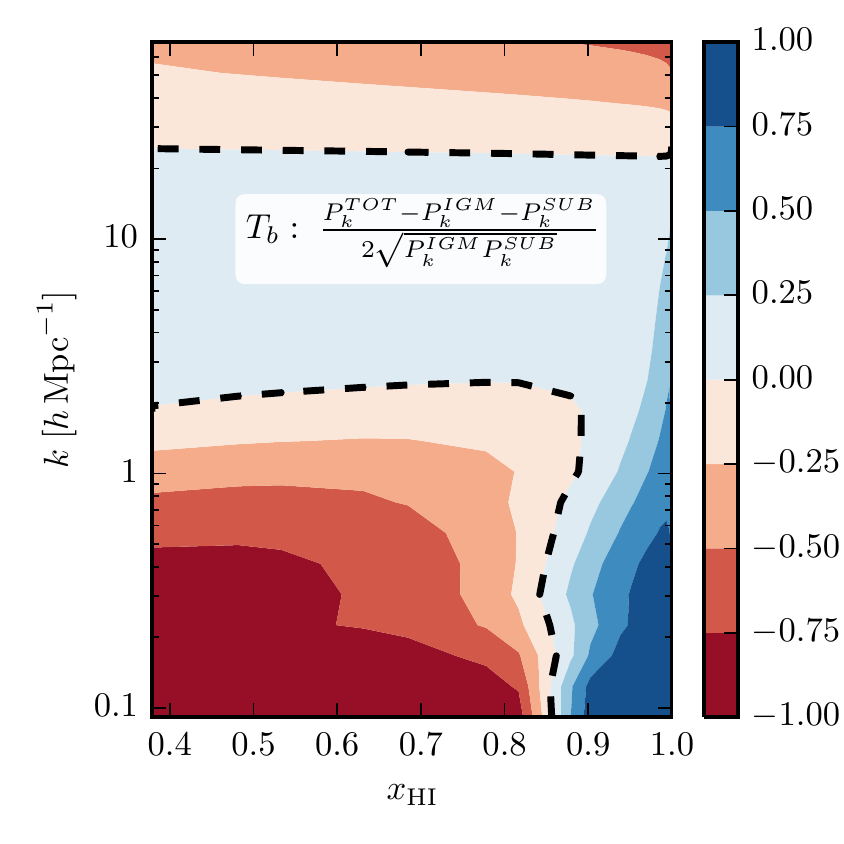}
\includegraphics[scale=1]{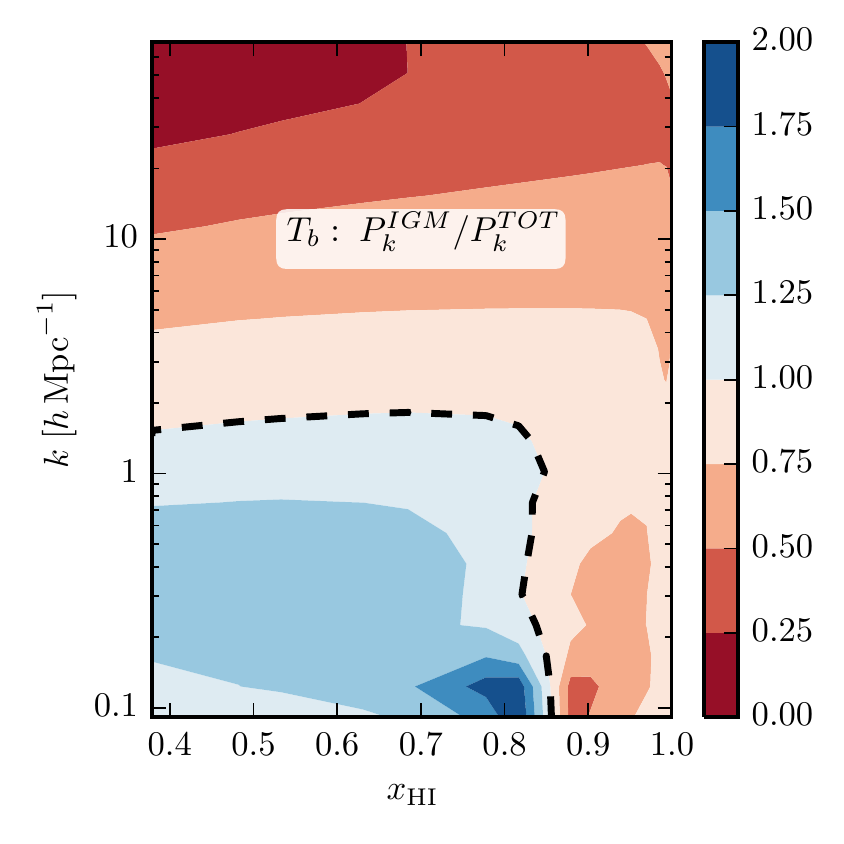}
\caption{Correlation coefficient $r(k)$ (left panel) and the ratio of the power spectrum for the neutral IGM component to the total power spectrum $P_k^{\rm IGM}/P_k^{\rm TOT}$ (right panel).}\label{fig:dev}
\end{figure*}

We can describe the discrepancy between the full power spectrum, $P_k^{\rm TOT}$, and the sum of the power spectra of individual components with the cross-correlation coefficient,
\begin{equation}
r(k) \equiv \dfrac{P_k^{\rm TOT}-P_k^{\rm IGM}-P_k^{\rm SUB}}{2\sqrt{P_k^{\rm IGM}P_k^{\rm SUB}}}.
\label{eq:r}
\end{equation}
The value of $r(k)$ quantifies the correlation between $\tbsub$ and $\tbigm$ fields. Negative sign corresponds to anti-correlation. The value of $r(k)$ is plotted in the left panel of Figure \ref{fig:Tbdiff}. In the right panel of Figure \ref{fig:dev} we plot the ratio $P_k^{\rm IGM}/P_k^{\rm TOT}$ that explicitly shows how far off $P_k^{\rm IGM}$ is from the true 21cm power spectrum $P_k^{\rm TOT}$.

During reionization the direct contribution from the neutral substructure, $P_k^{\rm SUB}$, remains small compared to the IGM contribution, $P_k^{\rm SUB} \sim (0.05-0.2) P_k^{\rm IGM}$. Hence, the effect of neutral substructure is entirely determined by the behavior of the correlation coefficient $r$. When $|r|\sim1$, neutral substructure contributes somewhere in the vicinity of $25-50\%$.

At highest redshifts ($z\ga10$, $x_\mathrm{HI}>0.85$) contributions from $\tbigm$ and $\tbsub$ are highly correlated, because the $\tbigm$ component closely traces the overall density field, and the $\tbsub$ component is its biased representation. Because in the beginning reionization proceeds inside-out, with time the $\tbigm$ field becomes less and less correlated with the $\tbsub$ field, until at $x_\mathrm{HI}\approx0.82$ the two fields become uncorrelated and the contribution of the $\tbsub$ field vanishes. At later times, i.e.\ during most of the reionization duration, the two field become strongly anti-correlated, and the contribution of the $\tbsub$ field again becomes significant (and negative), reducing the overall 21 cm power spectrum by up to 50\%.

Finally, after reionization is complete at $z\sim6$, the $\tbsub$ component becomes the only one. Overall, the $\tbsub$ component affects the 21 cm power spectrum in a non-trivial way at all redshifts and scales.

This behavior of the correlation coefficient $r$ can be understood analytically. Indeed, \citet{reisam:fzh04} derived an analytical expression for the the correlation function of quantity $\psi=x_\HI(1+\delta)$. On large scales, where the peculiar velocity effect becomes trivial, the actual brightness temperature of 21 cm emission is $\tb \propto \psi(T_S-T_{\rm CMB})/T_S$, where $T_S$ is the spin temperature. The second factor may, in principle, invalidate the \citet{reisam:fzh04} model; however, as we show in Appendix \S \ref{app:psi}, that factor is not important as soon as $\bar{x}_\HI$ deviates even slightly from unity, so the approximation $\tb\propto\psi$ is a very good one.

In that approximation \citet{reisam:fzh04} show that the correlation function for the $\tbigm$ component is (their Eq.\ 11)
\[
  \xi_{\psi} = \xi_{xx}(1+\xi_{\delta\delta}) + \bar{x}_\HI^2\xi_{\delta\delta} +\xi_{x\delta}(2\bar{x}_\HI+\xi_{x\delta}),
\]
where $\xi_{xx}\equiv\langle x_\HI(1)x_\HI(2)\rangle-\bar{x}_\HI^2$ is the auto-correlation function of the neutral hydrogen fraction, $\xi_{\delta\delta}$ is he auto-correlation function of density, and $\xi_{x\delta}$ is their cross-correlation. On large scales, where the correlation function is small, this equation can be simplified by retaining only linear in $\xi$ terms,
\begin{equation}
  \xi_{\psi} = \xi_{xx} + \bar{x}_\HI^2\xi_{\delta\delta} + 2\bar{x}_\HI\xi_{x\delta}.
  \label{eq:xipsi}
\end{equation}
With our definitions, we identify $\xi_\psi$ with $\xi_{\rm IGM}$. Including neutral substructure is equivalent (on large scales) to adding a biased density tracer to $\psi$,
\[
  \psi =  x_\HI(1+\delta) + b_f f_\HI (1+\delta),
\]
where $b_f$ is the filament bias factor and $f_\HI \ll 1$ is the fraction of neutral hydrogen remaining neutral within the ionized bubbles. With this definition for $\psi$, the total correlation function of neutral hydrogen becomes
\begin{equation}
  \xi_{\rm TOT} = \xi_{xx} + 2b_ff_\HI\xi_{x\delta} + b_f^2f_\HI^2\xi_{\delta\delta} + \bar{x}_\HI^2\xi_{\delta\delta} + 2\bar{x}_\HI(\xi_{x\delta}+b_ff_\HI\xi_{\delta\delta}).
  \label{eq:xitot}
\end{equation}
With $\xi_{\rm SUB} = b_f^2f_\HI^2\xi_{\delta\delta}$, we find
\[
  \xi_{\rm TOT} - \xi_{\rm IGM} - \xi_{\rm SUB} = 2 b_f f_\HI(\xi_{x\delta}+\bar{x}_\HI\xi_{\delta\delta}).
\]
Under the assumption that each halo of mass $M$ drives an ionized bubble of volume $V(M)$ around it, \citet{reisam:fzh04} derived an expressions for  $\xi_{x\delta} = -\bar{x}_\HI b_h b_i Q \xi_{\delta\delta}$, where $b_h = \int dM\, b(M)\, dn/dM$ is the average halo bias, $b_i = Q^{-1} \int dM\, b(M) V(M)\, dn/dM$ is the bubble-volume-weighted halo bias, and $Q= \int dM\, V(M)\, dn/dM$ is the IGM porosity.

An expression for $\xi_{xx}$ is not possible to derive in a closed form, but from the Schwartz inequality it can be written as $\xi_{xx} = w\bar{x}_\HI^2 b_h^2 b_i^2 Q^2  \xi_{\delta\delta}$, where $w\geq1$. With these expressions we find
\[
  \xi_{\rm TOT} - \xi_{\rm IGM} - \xi_{\rm SUB} = 2 b_f f_\HI \bar{x}_\HI\xi_{\delta\delta}(1-b_h b_i Q)
\]
and
\begin{eqnarray}
  r & = & \frac{2 b_f f_\HI \bar{x}_\HI\xi_{\delta\delta}(1-b_h b_i Q)}{2\left(\xi_{\rm IGM}\xi_{\rm SUB}\right)^{1/2}} \nonumber\\
    & = & \frac{1-b_hb_iQ}{\left(1+wb_h^2b_i^2Q^2-2b_hb_iQ\right)^{1/2}}.
  \label{eq:ran}
\end{eqnarray}
This expression for $r$ has the correct limiting behavior: $r\rightarrow 1$ for $Q\rightarrow 0$, $r$ changes sign at $\bar{x}_\HII \approx Q = 1/(b_hb_i) \approx 0.82$ if $b_hb_i\approx 5.5$, which is a very reasonable value, and $r\rightarrow -1$ for $Q\rightarrow \infty$, if the fudge factor $w$ goes to 1 in that limit.

\section{Conclusions}
\label{sec:conclusions}

In this study we check two commonly used (semi-) analytical approximations for modeling redshifted 21cm emission against the fully self-consistent numerical simulations. We highlight two effects that are important for analyzing observational data or generating mock data for the present and future 21 cm experiments.

First, we show that the global 21 cm signal is limited to within a $\pm(20-25)\dim{mK}$ range, in agreement with previous simulation studies \citep{ng:gs04} and in apparent disagreement with the analytical prediction of \citet{21cm:pl08}. That disagreement, however, is not substantive, but merely due to different assumptions about the relative strengths of the X-ray and Ly-$\alpha$ backgrounds. We expect our results, as based on fully self-consistent and observationally consistent modeling of the whole process of cosmic reionization, to be more accurate. This conclusion implies that observational efforts aimed at detecting the pre-reionization negative dip in the global signal will need to reach a factor of 4 to 5 higher sensitivity than is currently planned.

Second, we show that the signal from neutral hydrogen remaining in galaxies and filaments within ionized regions of the IGM should not be neglected. Even though this substructure gives a negligible contribution to the global 21 cm signal throughout the epoch of reionization, it changes the 21 cm power spectrum by 25-50\% at scales $k\sim 0.1-1h\dim{Mpc}^{-1}$. This range of scales is expected to be the most foreground free in the observations \citep{21cm:dbc10,21cm:twt12,21cm:dlw14,21cm:pld14}. However, it is unlikely that the first measurements will be capable of determining the shape of power spectrum with very high accuracy. Therefore, the first measurement is likely to be limited to the overall amplitude of the 21 cm signal only. The amplitude mostly depends on the heating mechanism, discussed in the first part of the paper, and also, as we show in the second part of the paper, on the often neglected substructure. Hence, models, which do not intrinsically account for the neutral substructure due to their limited resolution or adopted approximations, need to either add semi-neutral filaments explicitly in their modeled volumes or include corrections to the predicted 21 cm power spectrum in post-processing.

\begin{figure}
\includegraphics[scale=1.0]{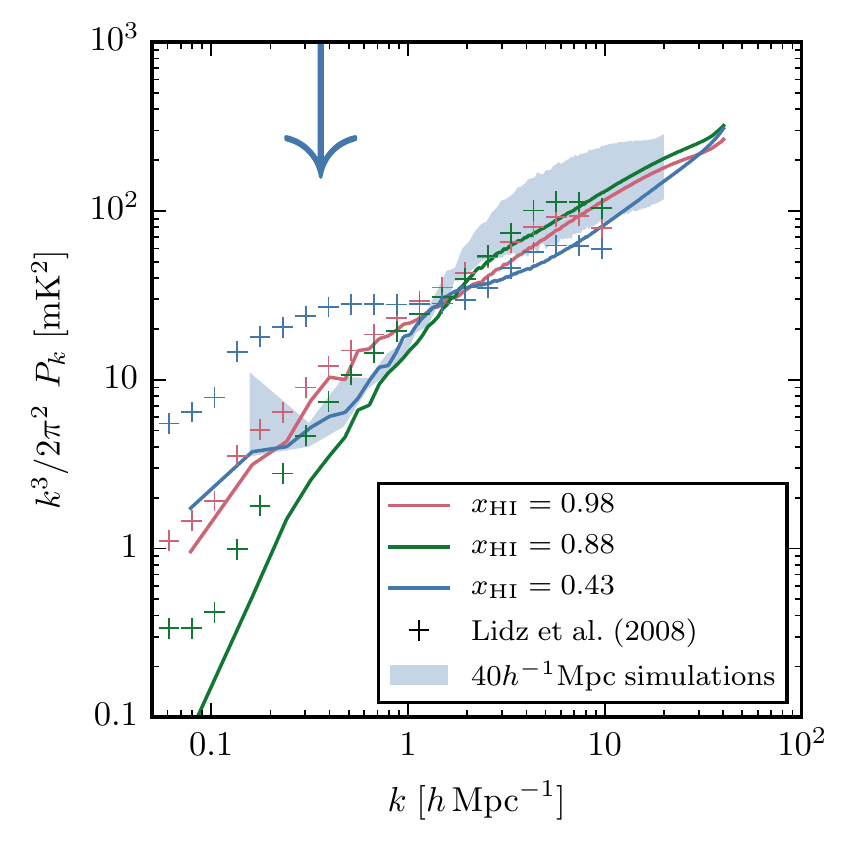}
\caption{Comparison of the 21 cm power spectrum from our simulations and from \citet{Lidz2008}. Solid curves show our largest $80h^{-1}\dim{Mpc}$ box at redshifts 8.8, 8.2 and 7.5. A shaded region corresponds to two standard deviation scatter between the three $40h^{-1}\dim{Mpc}$ runs, and can be used to estimate the numerical effect of the finite box size. Crosses trace the power spectrum calculated by \citet{Lidz2008} from their $130h^{-1}\dim{Mpc}$ run. Colors represent similar ionized fractions; however, actual redshifts are different -- 11.46, 8.8 and 7.3 correspondingly. The blue arrow shows the scale of the mean free path at redshift 7.5 (that corresponds to the blue lines and symbols) extrapolated using the fit by \citet{Songaila2010}. The discrepancy at low neutral fractions is larger than the substructure effect we discussed above.}\label{fig:lidz}
\end{figure}

However, other physical and numerical effects may influence theoretical predictions by much larger factors. In order to illustrate that, we compare in Figure \ref{fig:lidz} our predictions for the 21 cm power spectrum with the previous results from \citet{Lidz2008}, which is widely used as a target signal by experimental groups \citep{Bowman2013, 21cm:apz15}. For $\bar{x}_\HI<0.5$ the power at $k\sim0.1h\dim{Mpc}^{-1}$ in our run is about 5 times lower. One potential reason for the discrepancy is purely numerical - our largest box size is only $80h^{-1}\dim{Mpc}$, whereas \citet{Lidz2008} used a $130h^{-1}\dim{Mpc}$ box. In order to test the effect of the box size, we also show in Fig.\ \ref{fig:lidz} a blue band that encompasses $2-\sigma$ scatter estimated from 3 independent realizations of a $40h^{-1}\dim{Mpc}$ box. The box size seems to matter little for our simulations, but since we are not yet able to reach the $130h^{-1}\dim{Mpc}$ scale, a purely numerical source of the discrepancy can not yet be excluded.

Another potential source for the difference is actually physical: modeled physics is very different in our simulations and in \citet{Lidz2008}. In particular, CROC simulations fully account for the limited photon mean free path due to Lyman Limit systems, while \citet{Lidz2008} models allows photons to extend to arbitrary large distance. A limited photon mean free path may also limit the sizes of the largest bubbles (although the connection between the photon mean free path and the bubble size is not that direct, since bubbles around individual sources are clustered). The actual value of the mean free path due to Lyman Limit system in the CROC simulations is taken from \citet{Songaila2010} and is also shown in Fig.\ \ref{fig:lidz}. It is in the same range as the scales on which our results deviate significantly from  \citet{Lidz2008}, but whether this is indeed a reason for the discrepancy or just a coincidence is, of course, not possible to deduce from Fig.\ \ref{fig:lidz} alone, and would require a much more comprehensive and involved investigation, which goes well beyond the scope of this paper.

It appears, thus, that theoretical predictions for the 21 cm power spectrum still differ by factors as large as 5 even at the same $\bar{x}_\HI$ (and would differ even more at the same redshift, due to variations in the reionization history). Obviously, theorists still have some work to do before one can analyze and interpret any future observational measurement.

\acknowledgements

We thank Mat McQuinn and Jack Burns for valuable comments on the early draft of this paper. We also thank the anonymous referee for critical questions that made us to rethink and reinterpret our discussion in the original manuscript.

Fermilab is operated by Fermi Research Alliance, LLC, under Contract No.~DE-AC02-07CH11359 with the United States Department of Energy. This work was also supported in part by the NSF grant AST-1211190. Simulations used in this work have been performed on the National Energy Research Supercomputing Center (NERSC) supercomputers ``Hopper'' and ``Edison'' and on the Argonne Leadership Computing Facility supercomputer ``Mira''. An award of computer time was provided by the Innovative and Novel Computational Impact on Theory and Experiment (INCITE) program. This research used resources of the Argonne Leadership Computing Facility, which is a DOE Office of Science User Facility supported under Contract DE-AC02-06CH11357.

\appendix
\section{Spatial classification of regions}
\label{app:a}

\begin{figure}[b!]
\includegraphics[scale=1.0]{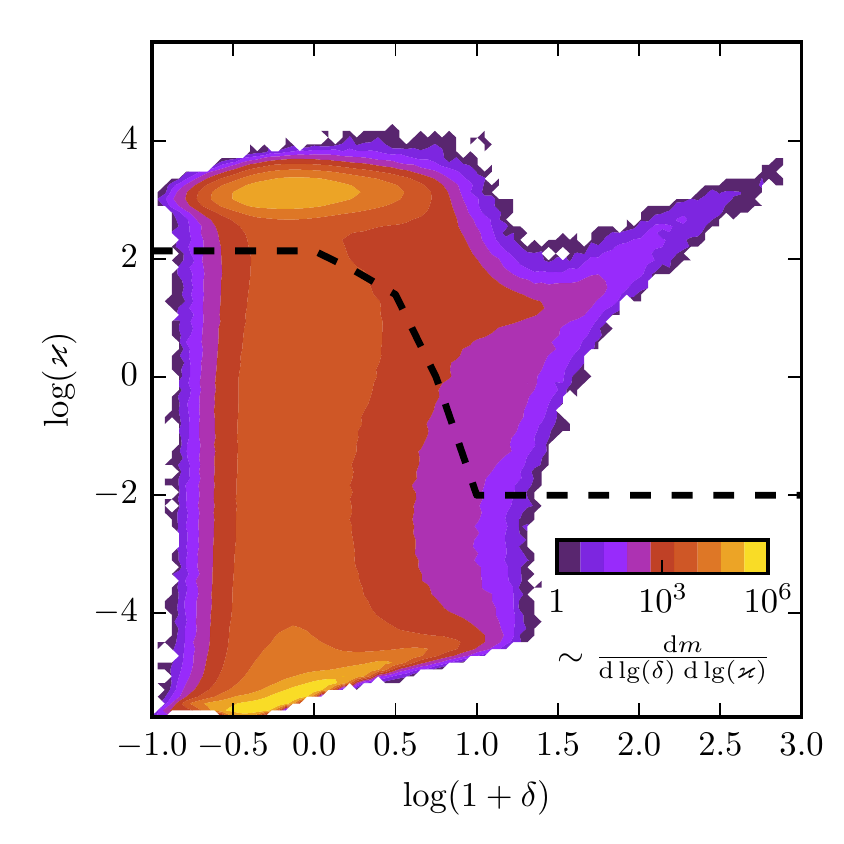}
\caption{\label{fig:phasespace}The phase space of the ionization state indicator, $\varkappa$ (Eq. \ref{eq:kappa}) and gas density at redshift 7.8 ($x_\mathrm{HI}=0.68$). The region above the dashed line corresponds to ionized bubbles and semi-neutral filaments ($\tbsub$); below the dashed line -- neutral IGM ($\tbigm$). }
\end{figure}

The ``ionization state indicator'' $\varkappa$ is defined as
\begin{equation}
\label{eq:kappa}
\varkappa = (1+\delta)\dfrac{x_{HII}^2}{x_{HI}},
\end{equation}
and its properties are described in details in \citet{ng:kg15}. In Figure \ref{fig:phasespace} the phase space diagram is plotted, along with the dashed line which is used to distinguish between ionized bubbles with embedded semi-neutral substructure ($\tbsub$) and yet-to-be-ionized IGM ($\tbigm$). Here we had a choice to assign ionized voids to either $\tbigm$ or $\tbsub$. Since the contribution from the ionized voids is minimal (they are more than 99.9\% ionized and underdense), it is not important whether they belong to one or the other. Therefore, we assign them to $\tbsub$ for simplicity.

In this paper we use downsampled snapshots of our simulations, rebinned into a $1024^3$ 2D$\times$1D (sky and frequency) grids; therefore, the information about the finest structure (galaxies and filaments) is degraded. That is why the phase plot in Figure \ref{fig:phasespace} looks less detailed, compared to those shown in \citet{ng:kg15}. Nevertheless, the resolution is still sufficient to observe main features, which allow to separate neutral IGM from neutral substructure.

\section{Approximation for the Brightness Temperature}
\label{app:psi}

\begin{figure*}
\includegraphics[scale=1.0]{Tb_deviations_r}
\includegraphics[scale=1.0]{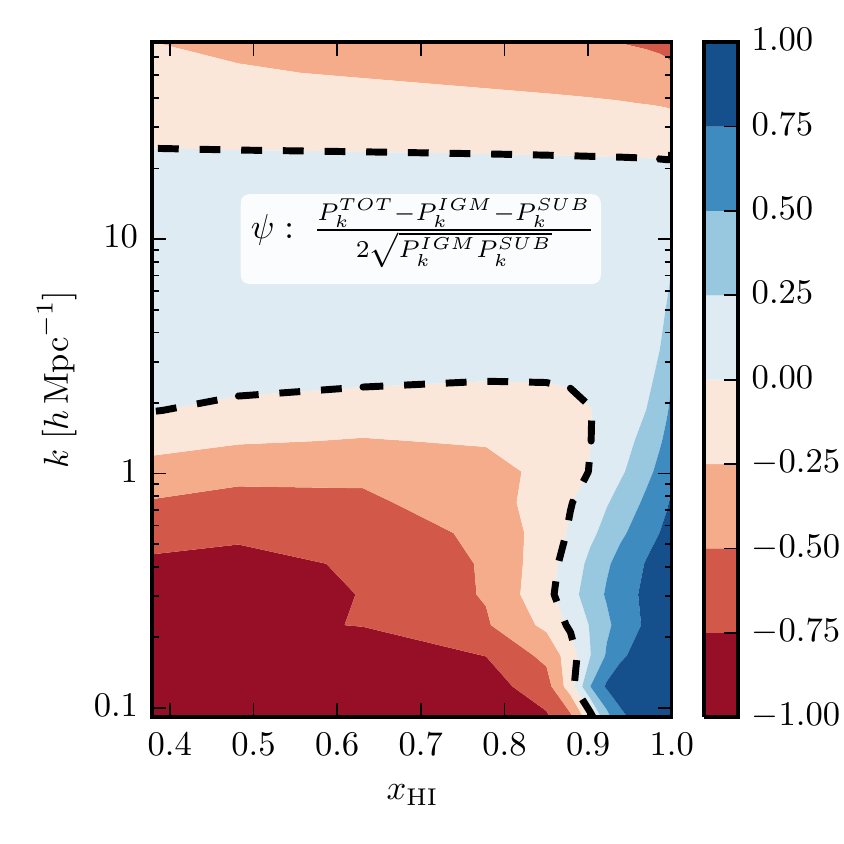}
\caption{Correlation coefficient $r$ (Eq.\ \ref{eq:r}) as a function of the wave number $k$ and redshift for the actual brightness temperature $\tb$ (left panel - identical to the left panel of Fig.\ \ref{fig:dev})) and for the approximate quantity $\psi=x_\HI(1+\delta)$, commonly used in analytical models (right panel). The approximate quantity $\psi$ provides a very good approximation to the actual brightness temperature. }\label{fig:Tbdiff}
\end{figure*}

On large scales, where peculiar velocity effects can be neglected, the brightness temperature of the 21 cm line is $\tb \propto x_\HI(1+\delta)(T_S-T_{\rm CMB})/T_S$. Analytical models, including the model of \citet{reisam:fzh04}, often operate with the quantity $\psi=x_\HI(1+\delta)$, ignoring the last temperature factor. In two panels of Figure \ref{fig:Tbdiff} we show the correlation coefficient $r$ (Eq.\ \ref{eq:r}) for both the actual brightness temperature $\tb$ and for $\psi$. The two panels are very similar, demonstrating that the spin temperature factor $(T_S-T_{\rm CMB})/T_S$ indeed can be neglected at $z \la 10$.

\bibliographystyle{apj}
\bibliography{refs,ng-bibs/igm,ng-bibs/self,ng-bibs/dsh,ng-bibs/rei,ng-bibs/21cm,ng-bibs/newrei,ng-bibs/reisam,ng-bibs/misc}

\begin{thebibliography}{35}
\expandafter\ifx\csname natexlab\endcsname\relax\def\natexlab#1{#1}\fi

\bibitem[{{Ali} {et~al.}(2015){Ali}, {Parsons}, {Zheng}, {Pober}, {Liu},
  {Aguirre}, {Bradley}, {Bernardi}, {Carilli}, {Cheng}, {DeBoer}, {Dexter},
  {Grobbelaar}, {Horrell}, {Jacobs}, {Klima}, {MacMahon}, {Maree}, {Moore},
  {Razavi}, {Stefan}, {Walbrugh}, \& {Walker}}]{21cm:apz15}
{Ali}, Z.~S., {Parsons}, A.~R., {Zheng}, H., {et~al.} 2015, \apj, 809, 61

\bibitem[{{Aubert} {et~al.}(2015){Aubert}, {Deparis}, \&
  {Ocvirk}}]{newrei:emma}
{Aubert}, D., {Deparis}, N., \& {Ocvirk}, P. 2015, \mnras, 454, 1012

\bibitem[{{Bowman} \& {Rogers}(2010)}]{21cm:br10}
{Bowman}, J.~D., \& {Rogers}, A.~E.~E. 2010, \nat, 468, 796

\bibitem[{{Bowman} {et~al.}(2013){Bowman}, {Cairns}, {Kaplan}, {Murphy},
  {Oberoi}, {Staveley-Smith}, {Arcus}, {Barnes}, {Bernardi}, {Briggs}, {Brown},
  {Bunton}, {Burgasser}, {Cappallo}, {Chatterjee}, {Corey}, {Coster},
  {Deshpande}, {deSouza}, {Emrich}, {Erickson}, {Goeke}, {Gaensler},
  {Greenhill}, {Harvey-Smith}, {Hazelton}, {Herne}, {Hewitt},
  {Johnston-Hollitt}, {Kasper}, {Kincaid}, {Koenig}, {Kratzenberg}, {Lonsdale},
  {Lynch}, {Matthews}, {McWhirter}, {Mitchell}, {Morales}, {Morgan}, {Ord},
  {Pathikulangara}, {Prabu}, {Remillard}, {Robishaw}, {Rogers}, {Roshi},
  {Salah}, {Sault}, {Shankar}, {Srivani}, {Stevens}, {Subrahmanyan}, {Tingay},
  {Wayth}, {Waterson}, {Webster}, {Whitney}, {Williams}, {Williams}, \&
  {Wyithe}}]{Bowman2013}
{Bowman}, J.~D., {Cairns}, I., {Kaplan}, D.~L., {et~al.} 2013, \pasa, 30, e031

\bibitem[{{Burns} {et~al.}(2012){Burns}, {Lazio}, {Bale}, {Bowman}, {Bradley},
  {Carilli}, {Furlanetto}, {Harker}, {Loeb}, \& {Pritchard}}]{21cm:blb12}
{Burns}, J.~O., {Lazio}, J., {Bale}, S., {et~al.} 2012, Advances in Space
  Research, 49, 433

\bibitem[{{Datta} {et~al.}(2010){Datta}, {Bowman}, \& {Carilli}}]{21cm:dbc10}
{Datta}, A., {Bowman}, J.~D., \& {Carilli}, C.~L. 2010, \apj, 724, 526

\bibitem[{{Dillon} {et~al.}(2014){Dillon}, {Liu}, {Williams}, {Hewitt},
  {Tegmark}, {Morgan}, {Levine}, {Morales}, {Tingay}, {Bernardi}, {Bowman},
  {Briggs}, {Cappallo}, {Emrich}, {Mitchell}, {Oberoi}, {Prabu}, {Wayth}, \&
  {Webster}}]{21cm:dlw14}
{Dillon}, J.~S., {Liu}, A., {Williams}, C.~L., {et~al.} 2014, \prd, 89, 023002

\bibitem[{{Dillon} {et~al.}(2015){Dillon}, {Neben}, {Hewitt}, {Tegmark},
  {Barry}, {Beardsley}, {Bowman}, {Briggs}, {Carroll}, {de Oliveira-Costa},
  {Ewall-Wice}, {Feng}, {Greenhill}, {Hazelton}, {Hernquist}, {Hurley-Walker},
  {Jacobs}, {Kim}, {Kittiwisit}, {Lenc}, {Line}, {Loeb}, {McKinley},
  {Mitchell}, {Morales}, {Offringa}, {Paul}, {Pindor}, {Pober}, {Procopio},
  {Riding}, {Sethi}, {Shankar}, {Subrahmanyan}, {Sullivan}, {Thyagarajan},
  {Tingay}, {Trott}, {Wayth}, {Webster}, {Wyithe}, {Bernardi}, {Cappallo},
  {Deshpande}, {Johnston-Hollitt}, {Kaplan}, {Lonsdale}, {McWhirter}, {Morgan},
  {Oberoi}, {Ord}, {Prabu}, {Srivani}, {Williams}, \& {Williams}}]{21cm:dnw15}
{Dillon}, J.~S., {Neben}, A.~R., {Hewitt}, J.~N., {et~al.} 2015, \prd, 91,
  123011

\bibitem[{{Duffy} {et~al.}(2014){Duffy}, {Wyithe}, {Mutch}, \&
  {Poole}}]{newrei:dwm15}
{Duffy}, A.~R., {Wyithe}, J.~S.~B., {Mutch}, S.~J., \& {Poole}, G.~B. 2014,
  \mnras, 443, 3435

\bibitem[{{Furlanetto} {et~al.}(2004){Furlanetto}, {Zaldarriaga}, \&
  {Hernquist}}]{reisam:fzh04}
{Furlanetto}, S.~R., {Zaldarriaga}, M., \& {Hernquist}, L. 2004, \apj, 613, 1

\bibitem[{{Gnedin}(2014)}]{ng:g14}
{Gnedin}, N.~Y. 2014, \apj, 793, 29

\bibitem[{{Gnedin} \& {Kaurov}(2014)}]{ng:gk14}
{Gnedin}, N.~Y., \& {Kaurov}, A.~A. 2014, \apj, 793, 30

\bibitem[{{Gnedin} \& {Shaver}(2004)}]{ng:gs04}
{Gnedin}, N.~Y., \& {Shaver}, P.~A. 2004, \apj, 608, 611

\bibitem[{{Greenhill} \& {Bernardi}(2012)}]{21cm:gb12}
{Greenhill}, L.~J., \& {Bernardi}, G. 2012, ArXiv e-prints

\bibitem[{{Jacobs} {et~al.}(2015){Jacobs}, {Pober}, {Parsons}, {Aguirre},
  {Ali}, {Bowman}, {Bradley}, {Carilli}, {DeBoer}, {Dexter}, {Gugliucci},
  {Klima}, {Liu}, {MacMahon}, {Manley}, {Moore}, {Stefan}, \&
  {Walbrugh}}]{21cm:jpp15}
{Jacobs}, D.~C., {Pober}, J.~C., {Parsons}, A.~R., {et~al.} 2015, \apj, 801, 51

\bibitem[{{Kaurov} \& {Gnedin}(2015)}]{ng:kg15}
{Kaurov}, A.~A., \& {Gnedin}, N.~Y. 2015, \apj, 810, 154

\bibitem[{{Leitherer} {et~al.}(1999){Leitherer}, {Schaerer}, {Goldader},
  {Delgado}, {Robert}, {Kune}, {de Mello}, {Devost}, \&
  {Heckman}}]{misc:lsgd99}
{Leitherer}, C., {Schaerer}, D., {Goldader}, J.~D., {et~al.} 1999, \apjs, 123,
  3

\bibitem[{{Lidz} {et~al.}(2008){Lidz}, {Zahn}, {McQuinn}, {Zaldarriaga}, \&
  {Hernquist}}]{Lidz2008}
{Lidz}, A., {Zahn}, O., {McQuinn}, M., {Zaldarriaga}, M., \& {Hernquist}, L.
  2008, \apj, 680, 962

\bibitem[{{McQuinn} \& {O'Leary}(2012)}]{21cm:mo12}
{McQuinn}, M., \& {O'Leary}, R.~M. 2012, \apj, 760, 3

\bibitem[{{Mesinger} {et~al.}(2011){Mesinger}, {Furlanetto}, \&
  {Cen}}]{21cm:mfc11}
{Mesinger}, A., {Furlanetto}, S., \& {Cen}, R. 2011, \mnras, 411, 955

\bibitem[{Miralda-Escud\'{e} {et~al.}(2000)Miralda-Escud\'{e}, Haehnelt, \&
  Rees}]{Miralda-Escude2000}
Miralda-Escud\'{e}, J., Haehnelt, M., \& Rees, M.~J. 2000, Astrophys. J., 530,
  1

\bibitem[{{Ocvirk} {et~al.}(2015){Ocvirk}, {Gillet}, {Shapiro}, {Aubert},
  {Iliev}, {Romain}, {Yepes}, {Choi}, {Sullivan}, {Knebe}, {Gottloeber},
  {D'Aloisio}, {Park}, \& {Hoffman}}]{newrei:dawn}
{Ocvirk}, P., {Gillet}, N., {Shapiro}, P., {et~al.} 2015, IAU General Assembly,
  22, 55292

\bibitem[{{O'Shea} {et~al.}(2015){O'Shea}, {Wise}, {Xu}, \&
  {Norman}}]{newrei:owx15}
{O'Shea}, B.~W., {Wise}, J.~H., {Xu}, H., \& {Norman}, M.~L. 2015, \apjl, 807,
  L12

\bibitem[{{Parsons} {et~al.}(2014){Parsons}, {Liu}, {Aguirre}, {Ali},
  {Bradley}, {Carilli}, {DeBoer}, {Dexter}, {Gugliucci}, {Jacobs}, {Klima},
  {MacMahon}, {Manley}, {Moore}, {Pober}, {Stefan}, \& {Walbrugh}}]{21cm:pla14}
{Parsons}, A.~R., {Liu}, A., {Aguirre}, J.~E., {et~al.} 2014, \apj, 788, 106

\bibitem[{{Patra} {et~al.}(2013){Patra}, {Subrahmanyan}, {Raghunathan}, \&
  {Udaya Shankar}}]{21cm:psr13}
{Patra}, N., {Subrahmanyan}, R., {Raghunathan}, A., \& {Udaya Shankar}, N.
  2013, Experimental Astronomy, 36, 319

\bibitem[{{Patra} {et~al.}(2015){Patra}, {Subrahmanyan}, {Sethi}, {Udaya
  Shankar}, \& {Raghunathan}}]{21cm:pss15}
{Patra}, N., {Subrahmanyan}, R., {Sethi}, S., {Udaya Shankar}, N., \&
  {Raghunathan}, A. 2015, \apj, 801, 138

\bibitem[{{Pober} {et~al.}(2014{\natexlab{a}}){Pober}, {Liu}, {Dillon},
  {Aguirre}, {Bowman}, {Bradley}, {Carilli}, {DeBoer}, {Hewitt}, {Jacobs},
  {McQuinn}, {Morales}, {Parsons}, {Tegmark}, \& {Werthimer}}]{21cm:hera}
{Pober}, J.~C., {Liu}, A., {Dillon}, J.~S., {et~al.} 2014{\natexlab{a}}, \apj,
  782, 66

\bibitem[{{Pober} {et~al.}(2014{\natexlab{b}}){Pober}, {Liu}, {Dillon},
  {Aguirre}, {Bowman}, {Bradley}, {Carilli}, {DeBoer}, {Hewitt}, {Jacobs},
  {McQuinn}, {Morales}, {Parsons}, {Tegmark}, \& {Werthimer}}]{21cm:pld14}
---. 2014{\natexlab{b}}, \apj, 782, 66

\bibitem[{{Pober} {et~al.}(2015){Pober}, {Ali}, {Parsons}, {McQuinn},
  {Aguirre}, {Bernardi}, {Bradley}, {Carilli}, {Cheng}, {DeBoer}, {Dexter},
  {Furlanetto}, {Grobbelaar}, {Horrell}, {Jacobs}, {Klima}, {Kohn}, {Liu},
  {MacMahon}, {Maree}, {Mesinger}, {Moore}, {Razavi-Ghods}, {Stefan},
  {Walbrugh}, {Walker}, \& {Zheng}}]{21cm:pap15}
{Pober}, J.~C., {Ali}, Z.~S., {Parsons}, A.~R., {et~al.} 2015, \apj, 809, 62

\bibitem[{{Pritchard} \& {Loeb}(2008)}]{21cm:pl08}
{Pritchard}, J.~R., \& {Loeb}, A. 2008, \prd, 78, 103511

\bibitem[{{Pritchard} \& {Loeb}(2012)}]{21cm:pl12}
---. 2012, Reports on Progress in Physics, 75, 086901

\bibitem[{{Sokolowski} {et~al.}(2015){Sokolowski}, {Tremblay}, {Wayth},
  {Tingay}, {Clarke}, {Roberts}, {Waterson}, {Ekers}, {Hall}, {Lewis},
  {Mossammaparast}, {Padhi}, {Schlagenhaufer}, {Sutinjo}, \&
  {Tickner}}]{21cm:stw15}
{Sokolowski}, M., {Tremblay}, S.~E., {Wayth}, R.~B., {et~al.} 2015, \pasa, 32,
  4

\bibitem[{{Songaila} \& {Cowie}(2010)}]{Songaila2010}
{Songaila}, A., \& {Cowie}, L.~L. 2010, \apj, 721, 1448

\bibitem[{{Trott} {et~al.}(2012){Trott}, {Wayth}, \& {Tingay}}]{21cm:twt12}
{Trott}, C.~M., {Wayth}, R.~B., \& {Tingay}, S.~J. 2012, \apj, 757, 101

\bibitem[{{Voytek} {et~al.}(2014){Voytek}, {Natarajan}, {J{\'a}uregui
  Garc{\'{\i}}a}, {Peterson}, \& {L{\'o}pez-Cruz}}]{21cm:vnj14}
{Voytek}, T.~C., {Natarajan}, A., {J{\'a}uregui Garc{\'{\i}}a}, J.~M.,
  {Peterson}, J.~B., \& {L{\'o}pez-Cruz}, O. 2014, \apjl, 782, L9

\end{thebibliography}

\end{document}